\documentclass[preprint,showpacs,preprintnumbers,amsmath,amssymb]{revtex4}
\usepackage{graphicx}
\usepackage{dcolumn}
\usepackage{bm}

\begin{document}


\title{On a possible observation of \lq\lq dihelion\rq\rq~in dissociation of relativistic $^9$C nuclei}

\author{D.~A.~Artemenkov}
   \affiliation{Joint Insitute for Nuclear Research, Dubna, Russia}
\author{N.~K.~Kornegrutsa}
   \affiliation{Joint Insitute for Nuclear Research, Dubna, Russia}
\author{D.~O.~Krivenkov}
   \affiliation{Joint Insitute for Nuclear Research, Dubna, Russia}  
\author{R.~Stanoeva}
   \affiliation{Joint Insitute for Nuclear Research, Dubna, Russia}
\author{P.~I.~Zarubin}
     \email{zarubin@lhe.jinr.ru}
     \homepage{http://becquerel.jinr.ru}
   \affiliation{Joint Insitute for Nuclear Research, Dubna, Russia}
   
\date{\today}

\begin{abstract}
\indent 
Nuclear track emulsion is exposed to 1.2~A~$^9$C~GeV nuclei. Pairs of 2$^3$He nuclei having unusually narrow opening angles are observed in channel $^9$C~$\rightarrow$~3$^3$He pointing to the possible 2$^3$He resonance near the production threshold.\par
\end{abstract}

\maketitle

\section{\label{sec:level1}Introduction}

\indent The BECQUEREL Collaboration explores the peripheral dissociation of light relativistic nuclei in nuclear emulsion \cite{web1}. The structural features of a projectile nucleus are most clearly manifested in the coherent dissociation of nuclei not accompanied by the formation of the target nucleus fragments and mesons (or in \lq\lq white\rq\rq~stars). Nucleon clustering in stable and radioactive Be, B, C and N isotopes are studied in this approach with a complete observation of the fragments at a record angular resolution \cite{Peresadko,Artemenkov1,Stanoeva,Kattabekov,Artemenkov2,Krivenkov,Rukoyatkin}. Macro photos of such interactions are assembled in the collection \cite{web1}.\par
\indent In particular, nuclear track emulsion is irradiated in a $^9$C nucleus beam produced at the JINR Nuclotron by the fragmentation of 1.2~A~GeV $^{12}$C nuclei \cite{Rukoyatkin}. Dominance of the $^9$C isotope in the secondary beam is confirmed by emulsion measurements of the ionization of the secondary beam nuclei, features of the charge topology of their fragmentation, as well as momentum measurements of the accompanying the $^3$He nuclei. As is established, the $^8$B~+~$p$, $^7$Be~+~2$p$, 2He~+~2$p$ and He~+~2H~+~2$p$ channels are leading ones in the $^9$C coherent dissociation \cite{Krivenkov}. The two latter are related to the $^7$Be core dissociation.\par
\indent In addition, population of the three $^3$He nucleus state is observed in the $^9$C coherent dissociation with 14\% probability. Its origin can be caused by a virtual regrouping of a neutron from the $^4$He cluster to form $^3$He clusters as well as the presence of the 3$^3$He component in the $^9$C ground state. In the second variant, the probability of the 3$^3$He ensemble production points to the significance of this deeply bound configuration in the wave function of the $^9$C ground state. The mechanism of the $^9$C coherent dissociation in the channels with nucleon separation and the 3$^3$He channel is a nuclear diffractive interaction which is established on the basis of measurements of the total transverse momentum transferred to the fragment ensemble.\par

\section{\label{sec:level2}Narrow  pairs in coherent dissociation  $^9$C~$\rightarrow$~3$^3$He}

\indent Few pairs of $^3$He nuclei with opening angles $\Theta_{2He}$ less than 10$^{-2}$ rad are detected in the $^9$C~$\rightarrow$~3$^3$He channel. The macro photo of one identified event is shown in figure 1. This observation indicates to a possible existence of a narrow 2$^3$He resonant state with the decay energy near the 2$^3$He mass threshold (or \lq\lq dihelion\rq\rq). In the same way the formation of $^8$Be nuclei is reliably manifested in the production of $^4$He pairs with extremely small opening angles in the relativistic dissociation of $^9$Be~$\rightarrow$~2$^4$He \cite{Artemenkov1} and $^{10}$C~$\rightarrow$~2$^4$He~+~2$p$ \cite{Stanoeva,Kattabekov}. Significant probability of coherent dissociation $^9$C~$\rightarrow$~3$^3$He makes it an efficient source for search for an analog of the unbound $^8$Be nucleus among the $^3$He pairs. In what follows, a rather unexpected and potentially important feature of the spectrum $\Theta_{2He}$ of $^3$He pairs produced in the dissociation of $^9$C \cite{Krivenkov} and $^8$B \cite{Rukoyatkin} nuclei, is given.\par 	

\begin{figure}
    \includegraphics[width=6in]{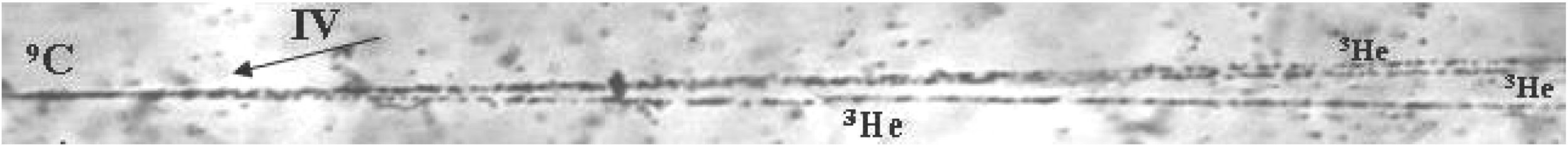}
    \caption{\label{Fig:1} Macro photo of \lq\lq white\rq\rq~star of $^9$C dissociation to 3$^3$He nuclei in nuclear track emulsion; the interaction vertex IV is shown by the arrow.}
    \end{figure}
	
\begin{figure}
    \includegraphics[width=5in]{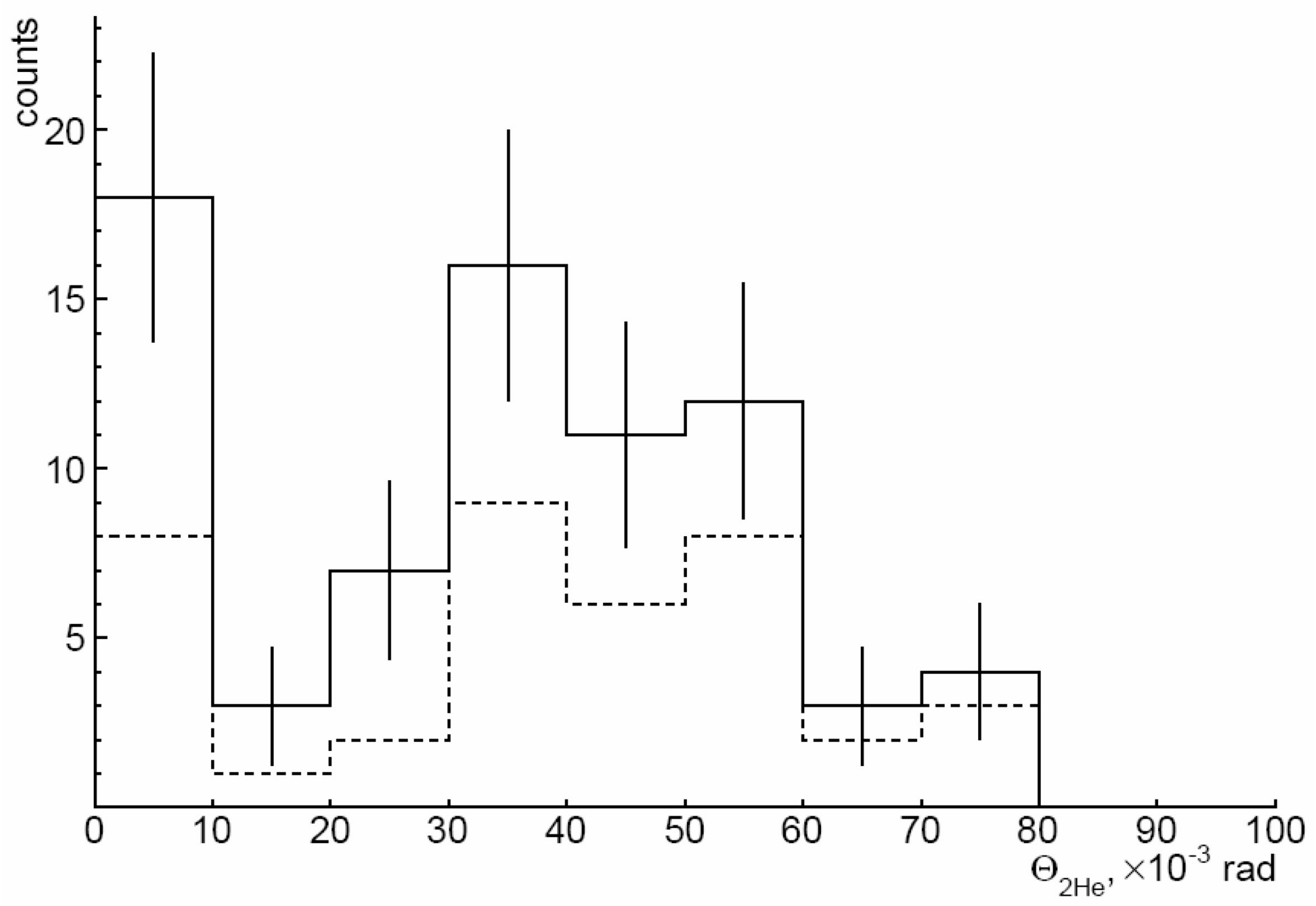}
    \caption{\label{Fig:2} Total distribution of opening angles $\Theta_{2He}$ between the relativistic He fragments in the \lq\lq white\rq\rq~stars $^9$C~$\rightarrow$~3$^3$He and in events $^8$B~$\rightarrow$~2He~+~H with the formation of target nucleus fragments or meson; dotted line indicates the \lq\lq white\rq\rq~stars contribution.}
    \end{figure}
	
\indent In figure 2 the dotted line shows the distribution $\Theta_{2He}$ for \lq\lq white\rq\rq~stars $^9$C~$\rightarrow$~3$^3$He. Its main part, corresponding to 30 pairs, is described by a Gaussian distribution with mean value $<\Theta_{2He}>$~=~(46~$\pm$~3)$\times$10$^{-3}$~rad (RMS 16$\times$10$^{-3}$~rad). In addition, thanks to excellent spatial resolution eight 2$^3$He pairs within $\Theta_{2He}~<~10^{-2}$~rad are reliably observed. These pairs form a special group with a mean value $<Q(2^3He)>$~=~(6~$\pm$~1)$\times$10$^{-3}$ rad and RMS 3$\times$10$^{-3}$ rad, which is obviously beyond the previous description. These values correspond to the average relative energy $<Q(2^3He)>$~=~(142~$\pm$~35)~keV (RMS 100~keV). The parameter $Q$(2$^3$He) was defined as the difference between the invariant mass of the pair and the double $^3$He mass assuming that the fragments conserve  the $^9$C momentum per nucleon.\par

\section{\label{sec:level3}Narrow  pairs in peripheral dissociation  $^8$B~$\rightarrow$~2He~+~H}

\indent Additional search for the resonance 2$^3$He is carried our in the events of 1.2~A~GeV peripheral dissociation $^8$B~$\rightarrow$~2He~+~H. In this case, inelastic interactions with target nucleus fragments or produced mesons are selected in order to enhance the \lq\lq dihelion\rq\rq~effect. This condition provides the selection of interactions with knocking of a neutron out of the $^4$He cluster in the $^8$B nucleus. Thus, the resulting distribution $\Theta_{2^3He}$ in the figure also includes a separate group of narrow pairs with $<\Theta_{2^3He}>$~=~(4.5~$\pm$~0.5)$\times$10$^{-3}$~rad (RMS 1.5$\times$10$^{-3}$~rad), corresponding to the case of \lq\lq white\rq\rq stars $^9$C~$\rightarrow$~3$^3$He.\par
\indent The total distribution for both the (figure 2) makes the indication to the existence of the 2$^3$He resonance more reliable. Moreover, the question arises about the nature of the broad peak with a maximum of $\Theta_{2^3He}$ about (40~$–$~50)$\times$10$^{-3}$ rad. It is possible that in this $\Theta_{2^3He}$ region the 2$^3$He system shows its similarity with the first excited 2$^+$ state of the $^8$Be nucleus.\par

\section{\label{sec:level4}Conclusions}

\indent Obviously, the \lq\lq dihelion\rq\rq~indication should be reviewed using a significantly larger statistics. The BECQUEREL Collaboration performed irradiation of nuclear track emulsion in a mixed beam of 1.2~A~GeV $^{12}$N, $^{10}$C and $^7$Be nuclei \cite{Kattabekov}. Thus, there are new opportunities with regard to the issue of \lq\lq dihelion\rq\rq~based on the analysis of the found 400 \lq\lq non-white\rq\rq~stars $^7$Be~$\rightarrow$~2$^3$He with knocking out of a neutron and the formation of fragments of target nuclei or mesons, as in the case of $^8$B~$\rightarrow$~2He~+~H.\par
\indent However, it is possible that the \lq\lq dihelion\rq\rq~formation is due to the presence of the 2$^3$He component in the $^9$C and $^8$B structures. In principle, in a lighter $^7$Be nucleus such a component can be suppressed, this means that the \lq\lq dihelion\rq\rq~formation can be suppressed as well. Therefore, it is important to search for the 2$^3$He resonance with high statistics exactly in low energy $^9$C and $^8$B beams. At the same time, pointing to the existence of \lq\lq dihelion\rq\rq, our observation motivates the search for a mirror state of a pair of nuclei $^3$H~-~\lq\lq ditriton\rq\rq.\par

\begin{acknowledgments}
\indent This work was supported by grants from 96-1596423, 02-02-164-12a, 03-02-16134, 03-02-17079, 04-02-17151. 04-02-16593, and 09-02-9126 CT-a RFBR. The authors thank S. P. Kharlamov (FIAN, Moscow), and A. I. Malakhov (JINR, Dubna) for discussions and friendly support.\par
\end{acknowledgments}

\end{document}